\documentclass[aps,amsmath,amssymb,prb,twocolumn,superscriptaddress]{revtex4}

\usepackage{bm}
\usepackage{epsfig}
\usepackage[usenames]{color}
\usepackage{graphicx}

\newcommand{\film}{{(s)}}
\newcommand{\substr}{{(p)}}

\newcommand{\nn}{\nonumber}

\begin{document}

\title{Emergence of superconducting textures in two dimensions}

\author{Andreas Glatz}
\affiliation{Materials Science Division, Argonne National Laboratory, Argonne, Illinois 60439, USA}

\author{Igor Aranson}
\affiliation{Materials Science Division, Argonne National Laboratory, Argonne, Illinois 60439, USA}

\author{Valerii Vinokur}
\affiliation{Materials Science Division, Argonne National Laboratory, Argonne, Illinois 60439, USA}

\author{Nikolay Chtchelkatchev}
\affiliation{L.D.\ Landau Institute for Theoretical Physics,
Russian Academy of Sciences, 117940 Moscow, Russia}
\affiliation{Department of Theoretical Physics, Moscow Institute
of Physics and Technology, 141700 Moscow, Russia}

\author{Tatyana Baturina}
\affiliation{Institute of Semiconductor Physics, 13 Lavrentjev Ave., Novosibirsk,
630090 Russia}

\date{\today}

\begin{abstract}
Self-organized regular patterns are ubiquitous in nature, and one of their most
celebrated manifestations is the Abrikosov vortex lattice\cite{Abrikosov}:
under an applied magnetic field, the homogeneous superconductivity
becomes unstable and cast itself into a regular texture of the ``normal'' filaments,
called Abrikosov vortices, immersed into a superconducting matrix.
Its prediction and the experimental discovery became a breakthrough
in our understanding of superconductivity and founded a new direction in physics.
Here we show that the interplay between the superconducting order parameter
and elastic fields\cite{Lasarev,Landauer,Olsen1,Olsen2,Shoenberg},
which are intimately connected to the very existence of the superconductivity
itself\cite{BCS}, can result in a novel superconducting state dual
to the Abrikosov state: a regular texture of superconducting islands.
The fact that both patterns emerge within the framework of the Ginzburg-Landau
description of superconductivity\cite{GL1950} indicates
that the formation of regular structures may be a generic
feature of any phase transition.
Emergence of superconducting
island arrays is not specific to the effect of the elastic forces, but can be caused
by any inherent mechanism generating long-range non-local interactions
in the Ginzburg-Landau functional, for example, by the Coulomb forces.
In particular, our findings suggest the formation of a superconducting island textures
as a scenario for a superconductor-to-insulator transition in thin films.

\end{abstract}

\maketitle

Sixty years ago the volume change accompanying the transition of a superconductor from
the normal to the superconducting state was first observed\cite{Lasarev}.
This discovery -- followed by finding the dependence of the superconducting critical
temperature, $T_{\mathrm c}$, on the isotopic mass\cite{isotop1,isotop2}
and the change in elastic constants at
the transition into the superconducting state\cite{Landauer,Olsen1,Olsen2} --
geared the line of research crowned eventually by the triumph of the
Bardeen-Cooper-Schrieffer (BCS) microscopic
theory of superconductivity,  which demonstrated an intimate connection between
superconductivity
and  elastic properties of the material\cite{Shoenberg}.
According to the BCS theory, $T_{\mathrm c}$ is related to elasticity via\cite{BCS}
       \begin{equation}
             k_{\mathrm {\scriptscriptstyle B}}T_{\mathrm c}=
             1.13\hbar\omega_{\mathrm{\scriptscriptstyle D}}
             \exp\left(-\frac{1}{\nu_0V}\right),
             \label{BCSeq}
       \end{equation}
where $\omega_{\mathrm{\scriptscriptstyle D}}$ is the Debye frequency, $\nu_0$
is the density of states at the Fermi level, and $V$ is the effective interaction strength between
electrons mediated by electron-phonon coupling.
All three parameters,
$\omega_{\mathrm{\scriptscriptstyle D}}, \nu_0$,
and $V$ in~(\ref{BCSeq}) depend on pressure and
comprise several microscopic effects including both the changes
in the electron and phonon spectrum, as well as structural transformations.
The link between elastic
properties and superconductivity
has been illustrated by numerous experiments revealing the
influence of an external pressure on $T_{\mathrm c}$
(see refs 11 and 12 for extensive reviews).

\begin{figure}
\includegraphics[width=0.95\columnwidth]{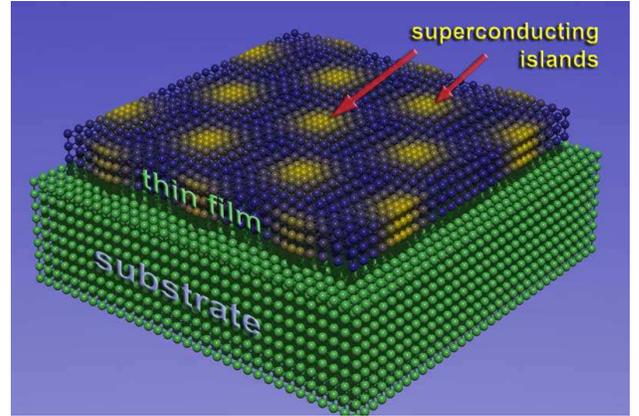}
\caption{{\bf The system.}
 A sketch of a ``soft'' superconducting
film deposited on a rigid substrate. Mechanical stresses induced by the substrate on the
film due to mismatch between the lattice constants of the film and of the
substrate, give rise to inhomogeneous superconducting state
which emerges in a form of a regular array
of separated superconducting islands.
}
\label{fig.model}
\end{figure}

Mechanical stresses, which affect superconductivity, are inherent to
thin films attached to rigid substrates\cite{Shih2009} (Fig.\,1).
Indeed, enforcing on the film its own lattice spacing, the substrate impels
\textit{internal} strains in the film.
On top of that, the rigid coupling between the
film and the substrate can give rise to
a peculiar scenario of the superconducting transition itself.
Upon cooling the film, the superconductivity
nucleates first in the regions where $T_{\mathrm c}$ is elevated by fluctuations.
Accordingly, mechanical properties, and, in particular,
the lattice spacing in these regions should have changed.
The rigid substrate, however, obstructs expansions (contractions) of the film
associated with the local onset of superconductivity.  As a result, additional local stresses
emerge, which, in their turn,
promote further growth (and the appearance of new) superconducting nuclei.
The  role of the substrate is thus twofold: first, it exerts
elastic forces in regions where superconducting droplets appear first and,
second, the substrate mediates the elastic
coupling between remote parts of the film, effectively
transforming local distortions into long-range elastic
coupling.
Appearing and evolving initial superconducting nuclei induce additional elastic
distortions.  As a result of this positive feedback between
the superconducting order parameter and mechanical stress, a spatial
instability of the superconducting state develops forming eventually
a periodic pattern of superconducting islands (see Fig.\,1).

\begin{figure*}
\includegraphics[width=0.8\textwidth]{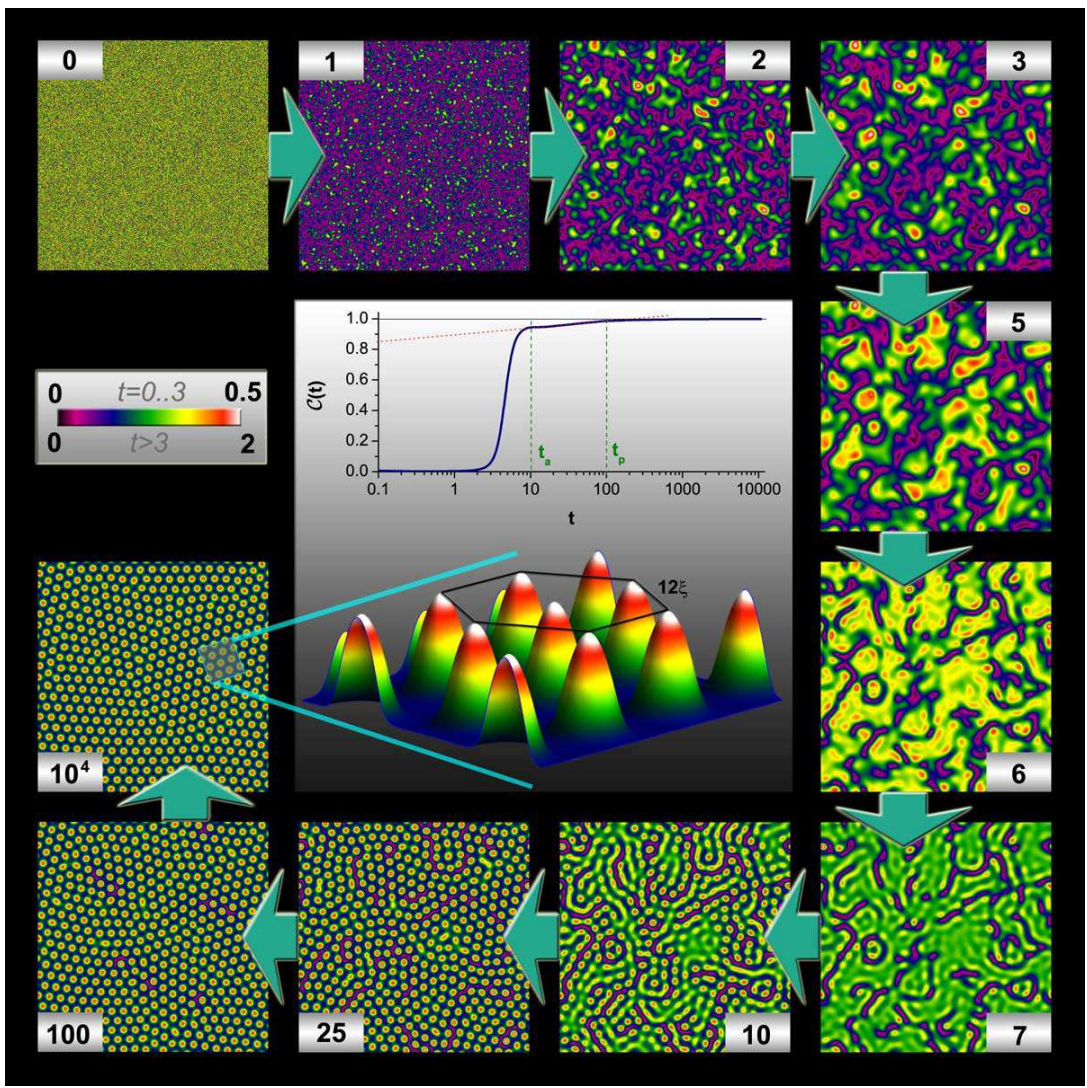}
\caption{{\bf Formation of the island texture.}
The framing sequence of snapshots shows the temporal evolution of the spatial distribution of the
amplitude of the superconducting order parameter.
The time is measured in units of the Ginzburg-Landau time, which for temperature $T=0.8T_c$
can be estimated as $\tau_{\mathrm{\scriptscriptstyle GL}}\simeq 10^{-11}$\,seconds
for $T_c\approx 1$K. The first frame at time $t=0$ shows an initial random configuration
of the order parameter. One can distinguish three stages of the evolution:
(i) emergence of an amorphous structure of islands (for $t<t_a\sim 10$);
(ii) formation of a polycrystalline islands texture  (for $t_a<t<t_p\sim 100$)
and (iii) relaxation to a long-range ordered island lattice (for $t>10^4$).
The color bar for the amplitude of the order parameter is presented beneath the initial
frame (note the change of the scale for $t>3$).
The upper part of the central panel shows the time
evolution of the normalized order parameter correlation function,
${\cal C}(t)={\cal N}^{-1}\sum_{k}|\psi_k|^2$, vs $t$ on a logarithmic scale
($\psi_k$ denote the Fourier components of the order parameter and the normalization factor
${\cal N}$ is chosen such that ${\cal C}=1$ when the island texture is fully periodic).
At intermediate times between $t_{\mathrm a}$ and $t_{\mathrm p}$ the correlation
function shows transient logarithmic behavior (highlighted by the straight line)
and exhibits for $t>t_p$ a slow convergence to unity.
The lower part shows a perspective view of the
height profile of the amplitude of the  order parameter corresponding to a small region of the
perfect lattice appearing at the final stage of solution of the TDGL equation.
The simulations were done with the coupling constant of
$U_0=2.23U_c$ and the thickness of the film of $0.8\xi$,
where $\xi$ is the superconducting coherence length at zero temperature.
Defining an ``island" as an area within which
the amplitude of the order parameter exceeds half of its maximal value,
we find their size to be about  $2.5\xi$ and the distance between
the centers of the islands to be $12\xi$.
}
\label{fig.islands3D}
\end{figure*}

Our starting point is the time-dependent Ginzburg-Landau (TDGL)
equation\cite{GL1950,kopnin_book,aranson_2002RMP} combined
with elastic stress balance equations (see Methods).
This model offers a comprehensive description
of formation and temporal evolution of the superconducting nuclei
towards the final equilibrium configuration of the order parameter.
We choose a random initial
configuration for the order parameter which
mimics the random distribution of the superconducting nuclei
expected to form in a homogeneously
disordered film upon cooling it down to $T_{\mathrm c}$.
However, the final equilibrium island pattern does not depend
on the particular choice of the initial configuration.
The advantage of this approach is that all the richness and
complexity of the microscopic elasticity-superconductivity
interrelations is accounted for by the phenomenological
coupling constant
$U_0=3\alpha_{\mathrm {\scriptscriptstyle L}}K(\partial T_{\mathrm c}/\partial p)$,
where $p$ is
the pressure, $K$ is the bulk elastic modulus, and
$\alpha_{\mathrm {\scriptscriptstyle L}}=(1/L)(\partial L/\partial T)$ is
the linear expansion parameter
(with $L$ being the linear dimension of the film).
These quantities can be inferred from experiments and contain
all the information about the electronic degrees of freedom and
the lattice excitation spectrum\cite{note2}.

\begin{figure*}
\includegraphics[width=0.8\textwidth]{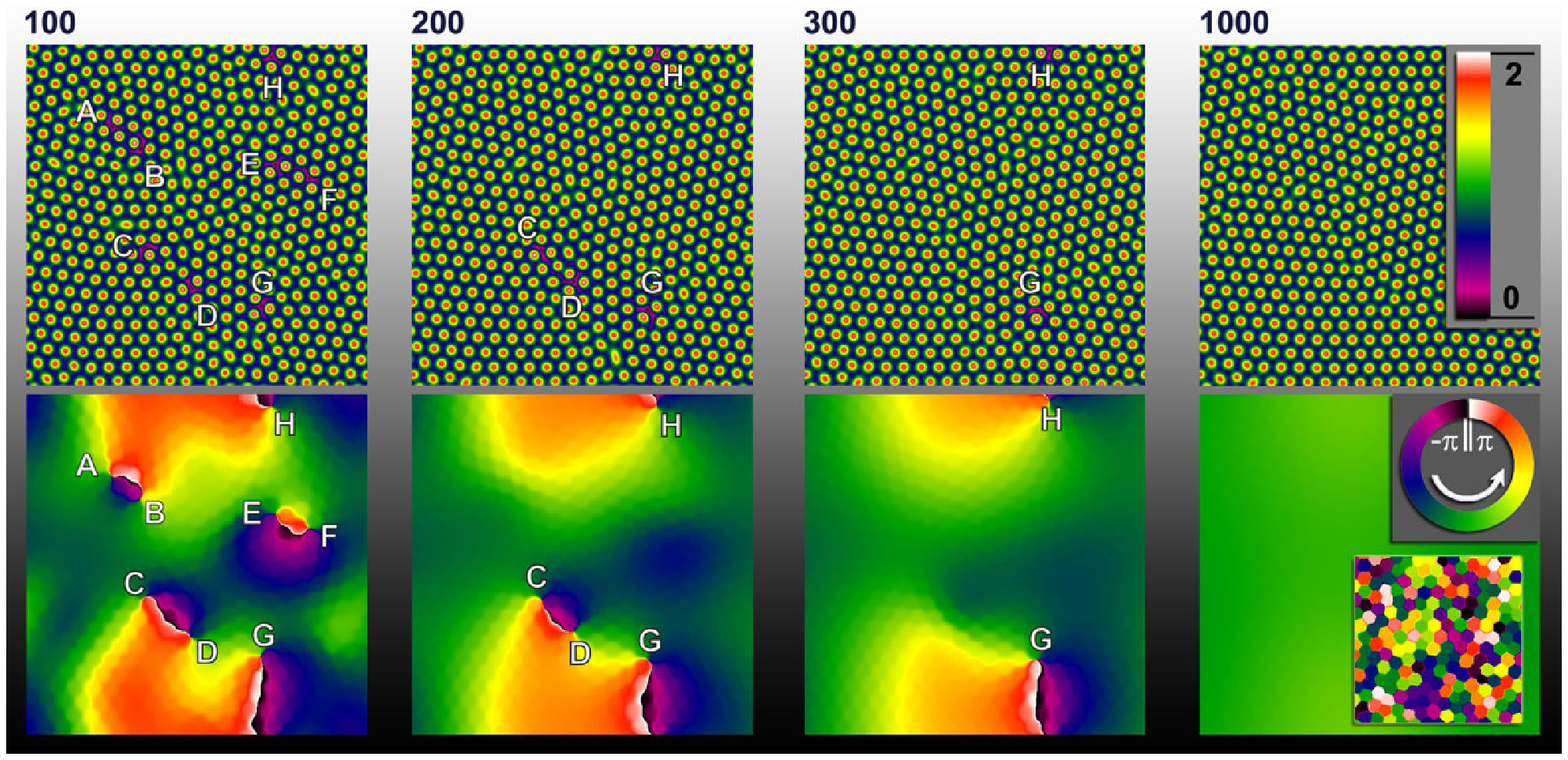}
\caption{{\bf Temporal evolution of the amplitude and phase of the order parameter.} Four
sequential snapshots of the amplitude and phase of the order parameter taken at
times 100, 200, 300, and 1000 measured in units of
$\tau_{\mathrm{\scriptscriptstyle GL}}$ are presented (the same values of the system parameters
as in Figure 2 were used).
The first three frames correspond to nonequilibrium states characterized by an inhomogeneous
phase distribution; in this stage the formation of a regular
texture of the order parameter amplitude
goes along with the recombination of vortex-antivortex pairs.
The first frame displays four sets of vortex-antivortex pairs with
the endpoints \textbf{A-B}, \textbf{C-D}, \textbf{E-F}, and \textbf{G-H}.
In the corresponding phase frames these endpoints confine the line of $2\pi$ phase jumps
visible as sharp color change from white to black (the periodic boundary conditions were used,
such that the line \textbf{G-H} goes over the upper- and lower edges of the frame).
The phase cut lines \textbf{A-B} and \textbf{E-F} disappear at $t=200$,
and at $t=300$ only the \textbf{G-H} line remains.
In the nonequilibrium states, the hexagon-like structure of the phase
distribution reflecting that on the way to equilibrium the phase at the islands
are different is clearly distinguishable (each hexagon represents the phase of an island).
At $t=1000$ the phase becomes homogeneous
across the sample and a global phase-coherent superconducting state establishes. However, for a different set of parameters the phase equilibration might be very slow or even stopped in the presence of impurities or dissipation, resulting in a random pattern of phases for each island, shown in the inset of the last phase-frame.}
\label{fig.islands-phase}
\end{figure*}

In the absence of elastic interactions due to connection to a substrate, one has the standard
equilibrium solution of the conventional TDGL equation:
a spatially uniform order parameter $\psi=\psi_0$ describing
the homogeneous superconducting state at $T<T_{\mathrm c}$, and
$\psi=0$ at $T>T_{\mathrm c}$.
The non-local elastic interaction, $U(\mathbf{r-r^{\prime}})$,
coupling the values of the superconducting order parameter at
different points ${\mathbf r}$ and $\mathbf{r^{\prime}}$ of the film
[see Eq.\,(3) of Methods] distorts $\psi_0$ and can
give rise to an instability of the uniform solution resulting
in the formation of a regular island texture for certain values of the elastic parameters.
To investigate this process
we solve the coupled TDGL and elasticity equations by numerical
integration using a quasi-spectral technique.
Figure 2 displays a sequence of snapshots for the temporal
development of the spatial structure of the amplitude of the order parameter.
The time is measured in the units of the Ginzburg-Landau time
$\tau_{\mathrm{\scriptscriptstyle GL}}$, with
$\tau_{\mathrm{\scriptscriptstyle GL}}={\pi\hbar}/
[8k_{\mathrm{\scriptscriptstyle B}}(T_{\mathrm c}-T)]$.
Starting from the random order parameter configuration, the system
evolves through three clearly distinguishable major stages:
(i) initial amplification of small fluctuations and emergence of an amorphous structure of islands;
(ii)  appearance of a polycrystalline configuration of well
separated superconducting islands;
and (iii) slow relaxation of polycrystalline structure to a regular island lattice.
The evolution of the modulus of the order parameter is quantified by the correlation function
${\cal C}(t)={\cal N}^{-1}\sum_{k}|\psi_k|^2$, where $\psi_k$ denote the Fourier components
of the order parameter, and the normalization factor ${\cal N}$ is chosen such
that ${\cal C}=1$ when the island texture is fully periodic.

The first stage is relatively fast: establishing of an amorphous island
pattern takes only about 10$\tau_{\mathrm{\scriptscriptstyle GL}}$, while
achieving a polycrystalline structure requires a 10 times longer period.
In this intermediate time scale ${\cal C}(t)$ evolves logarithmically towards
the polycrystalline state. In the final stage, for $t>t_{\mathrm p}$, see Fig.\,2,
where the island polycrystal relaxes towards the
regular lattice, the correlation function of the modulus of the order parameter
ceases to be an indicative characteristic quantity.
At longest time-scales it is rather the temporal development of the spatial distribution
of the \textit{phase} of the order parameter that characterizes the evolution of the system.
For the chosen material constants the final state is a superconducting state
with the uniform phase distribution corresponding to a long-ranged phase coherence.
Figure 3 shows that the macroscopic phase-coherent state
(which appears as a uniformly colored frame) is achieved via the
motion and recombination of vortex-antivortex pairs which are initially present
\textit{en mass} in the system due to coalescence of the independent
superconducting nuclei in the first stage of the order parameter evolution.
The vortex-antivortex pairs are clearly displayed by the ``phase-cut" lines that
appear as sharp color jumps from black to white.  In the same time interval
one distinguishes a well pronounced hexagon substructure which shows that different
islands have different phases.
The global phase coherence and therefore the superconducting
state get established by the time of order
$10^3 \tau_{\mathrm{\scriptscriptstyle GL}}$, which is by an order of
magnitude longer than the
time for establishing a robust distribution of the amplitude of the order parameter.

\begin{figure}
\includegraphics[width=0.98\columnwidth]{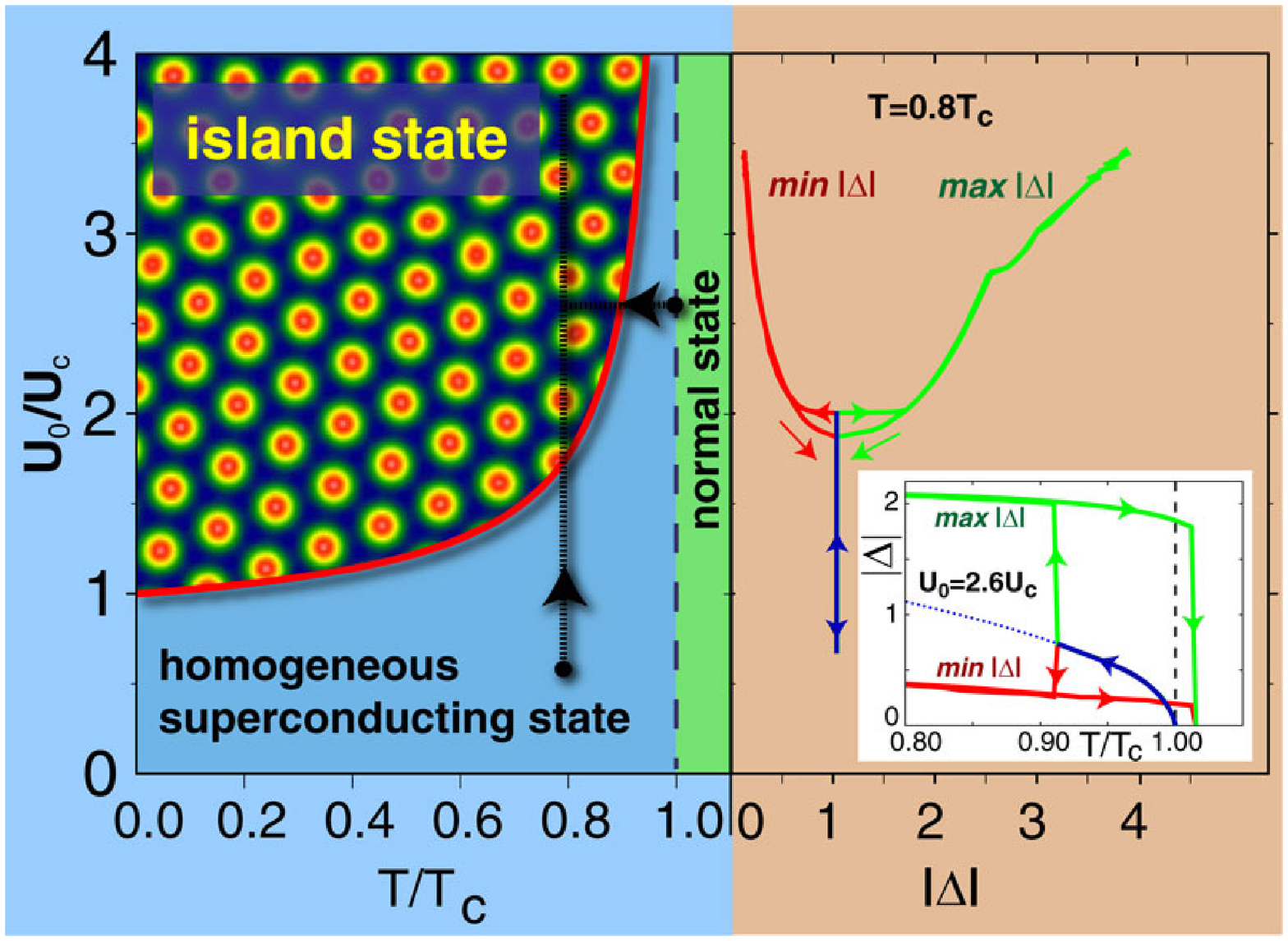}
\caption{{\bf Phase diagram and hysteretic onset of the island texture.}
The left panel shows the phases in the temperature
$T$ - coupling constant, $U_0$, plane.
Above $T_{\mathrm c}$ the film is in the normal state,
whereas below $T_{\mathrm c}$ the film can be either a homogeneous superconductor or a
textured superconductor consisting of isolated islands.
The phase boundary is determined by
the stability condition of the TGLE (see text). The right panel
shows the amplitude of the order parameter $\Delta$ as function of the
coupling constant $U_0$. The onset of the island texture (crossing
the phase boundary) is marked by the bifurcation point where
$\Delta_{\mathrm {max}}$ and $\Delta_{\mathrm {min}}$ start to diverge.
The observed hysteretic behavior -- the island structure forms,
upon adiabatically increasing the coupling, at larger value of $U_0$ than the reverse
transition, upon decreasing $U_0$, to the spatially homogeneous order parameter
state -- indicates
that the formation of the island texture occurs via a first order transition.
The path of change of $U_0$ is indicated by the vertical arrow in the left panel.
The inset in the right panel shows the hysteretic effect corresponding
sweeping temperature at fixed $U_0=2.6U_{\mathrm c}$, first,
from $T_{\mathrm c}$ down to $0.8T_{\mathrm c}$, and then heating
up (the process shown by the horizontal dotted line in the left panel).
It is noteworthy that the island structure persists even upon heating the
system above $T_{\mathrm c}$ of the film material, due to the
increased local $T_{\mathrm c}$ within the islands.
The dotted line indicates the usual $\sqrt{1-T/T_{\mathrm c}}$
behavior of the order parameter without elastic interactions.}
\label{fig.phases}
\end{figure}

It is important that, while being a general phenomenon, the formation
of an island texture requires the elastic coupling to be strong enough.
Namely, the texture appears under the condition that
the coupling constant $U_0$ exceeds
some temperature dependent critical value $U_{\mathrm c}(T)$.
The function
$U_{\mathrm c}(T)$ is determined by means of linear stability analysis of the TDGL.
Writing down the Fourier transform of the elastic interaction term
$U(\mathbf{r})$ as $\tilde U(q)=U_0{\cal K}(q)$, where ${\cal K}(q)$ is the scale free elastic
kernel in the momentum space and $\gamma$ is the diffusion coefficient
in the Ginzburg-Landau equation (see Method),
one finds that the island state forms if
$(U_0{\cal K}_{\mathrm{\scriptscriptstyle M}}-1)(1-T)-\gamma
q_{\mathrm{\scriptscriptstyle M}}^2/9.38\geq 0$,
where ${\cal K}_{\mathrm{\scriptscriptstyle M}}$
is the maximum value of ${\cal K}(q)$ at $q=q_{\mathrm{\scriptscriptstyle M}}$.
This gives $U_{\mathrm c}(T)={\cal K}_{\mathrm{\scriptscriptstyle M}}^{-1}+(\gamma
q_{\mathrm{\scriptscriptstyle M}}^2/9.38{\cal K}_{\mathrm{\scriptscriptstyle M}})(1-T/T_{\mathrm
c})^{-1}$.  The lower bound for the coupling constant at which the
homogeneous state becomes unstable and the island texture can appear is
$U_{\mathrm c}=(\gamma q_{\mathrm{\scriptscriptstyle M}}^2/9.38+1)/
{\cal K}_{\mathrm{\scriptscriptstyle M}}$.
The left panel of the Figure~4
shows the phase diagram in the reduced
$T/T_{\mathrm c}$-$U_0/U_{\mathrm c}$ coordinates for the chosen system parameters which are
given in the Methods.
The right panel of Fig.\,4 shows the behavior
of the amplitude of the order parameter $|\Delta|$ at the isotherm $T=0.8T_{\mathrm c}$ as
function of the coupling constant $U_0$. The protocol for finding this
dependence is as follows:
We start with a coupling constant $U_0<U_{\mathrm c}$ and an initial
random distribution of the order parameter, and let the system evolve
until a stationary distribution of $\Delta$ is achieved.
After that $U_0$ is increased
and the system evolves again until a new stationary state is achieved
(adiabatic increase), and so forth.
Upon crossing the phase boundary a bifurcation in the $|\Delta(U_0)|$-dependence
corresponding to the islands formation occurs and a \textit{finite} difference between
the maximal and minimal values of the amplitude of the order parameter appears.
In other words $|\Delta(U_0)|$ shows finite jumps to the $|\Delta_{\mathrm{min}}(U_0)|$
and $|\Delta_{\mathrm{max}}(U_0)|$ branches at some $U_0$ appreciably
beyond the phase boundary.  On the descending part of the cycle,
$|\Delta_{\mathrm{min}}(U_0)|$ and $|\Delta_{\mathrm{max}}(U_0)|$
merge at smaller value of  $U_0$ closer to the phase
boundary, i.e. at smaller $U_0$.  The change of $U_0$ is indicated by the
vertical arrow in the left panel of Fig.\,4.
Shown in the right panel inset is the hysteretic behavior of the order
parameter corresponding
to sweeping the temperature across the phase transition line.
The protocol is nearly the same as for the coupling constant variation scheme;
the temperature is first decreased from $T=T_{\mathrm c}$ down to $T=0.8T_{\mathrm c}$
and then increased back at the fixed coupling constant $U_0=2.6U_{\mathrm c}$.
The temperature sweeping is indicated by the horizontal arrow in the left panel of Fig.~4.
Note that the island structure persists even as the system is heated above
$T_{\mathrm c}$; this is the manifestation of the positive feedback between the
superconductivity and elasticity in this films bonded to the substrate giving rise to
a local increase in superconducting transition temperature within the islands.
The observed hysteretic behaviors indicate that the transition between
the homogeneous superconducting and the island
texture states is of the first order.

The pattern formation and, in particular, the spontaneous emergence of
electronic nanometer-scale structures, due to the existence of competing
states is ubiquitous in nature and is found in a wealth of complex systems
and physical phenomena ranging from magnetism,
superconductivity and superfluidity
to liquid crystals and biological dynamics\cite{aranson_2002RMP,dagotto-s05}.
The Ginzburg-Landau equation, one of the most universal models in
theoretical physics, offers an adequate quantitative description of the pattern formation.
Here we have revealed the instability and emergence of an electronic
textures, due to coupling of the superconducting
order parameter with elastic stress fields, generating a non-local term
in the Ginzburg-Landau functional.
It is noteworthy that the general form of the
nonlocality in the Ginzburg-Landau equation, induced by the interplay between the
local nonlinearity and the nonlocal couplings, is universal and can result from
various physical mechanisms.
In particular, while the results presented here
are stable with respect to weak disorder, we expect
that the interplay of strong disorder with
Coulomb interactions will give an essential contribution to the nonlocal coupling
in the Ginzburg-Landau functional.
Note, in this connection the interesting discussion
of the effects of disorder leading to a spatially inhomogeneous distribution of
the order parameter\cite{ghosal+prl998,ghosal+prb2001,Dubi_Nature,Imry2008,Sacepe08}.
We thus conclude that a consistent consideration of the
superconductivity in thin films has to include the effects of
the interplay between strong disorder with Coulomb interactions and elasticity.

The observed island texture
is an array of granules of a ``good" superconductor immersed
into a sea of a ``bad" one with much smaller amplitude of the order parameter and
is equivalent to an array of small superconducting droplets connected via weak
links, i.e., a Josephson junction array (JJA).
In our simulations, where thermal and
quantum fluctuations are absent, the Josephson coupling, however weak,
eventually establishes a global phase coherence and thus
global superconductivity.
The fluctuations, dissipation, and disorder
can break down this phase uniformity: the intermediate stages
of phase distributions of our simulations represent
the final states of the``real world'' systems.
Thus the possible eventual configuration   of the order parameter phase
(see the ``Gaudi-mosaic'' phase distribution in
the inset of the last phase-frame of the Fig. 3)
can correspond even the insulating state.
This is in accord with the general result that JJAs turn insulators if
the Josephson coupling between the superconducting islands
becomes sufficiently weak\cite{Fazio1991}.
Thus our finding that the transition into a superconducting state
in two dimensional superconductors elastically tied to a substrate
can occur via the formation of nanoscale superconducting island pattern
offers a possible scenario for
the superconductor-to-insulator transition observed in various
thin superconducting films\cite{Be1,Be2,InO1,InO2,TiN1,TiN2}.
This supports the conjecture\cite{FVB,vinokur+n08} that the
formation of a regular array of superconducting droplets is an inherent
property of the critical region of the superconductor-to-insulator
transition.

\acknowledgements

We thank Yuri Galperin and David Hinks for
useful discussions. This work was supported by the
U.S. Department of Energy Office of Science under the Contract No. DE-AC02-06CH11357,
by the Program ``Quantum macrophysics'' of the Russian Academy of Sciences,
and by the Russian Foundation for Basic Research (Grant Nos. 09-02-01205 and 09-0212206).

\appendix

\section{Addendum}

\subsection{Ginzburg-Landau equation.}
The two-dimensional time-dependent Ginzburg-Landau equation (TGLE)
describing the behavior of the film is written, in dimensionless variables as:
\begin{equation}\label{eq.tgle}
\partial _{t}\psi =\alpha\psi -\beta |\psi |^{2}\psi +\gamma \mathbf{\nabla }
^{2}\psi-\delta |\psi |^{4}\psi\,,
\end{equation}
where $\psi=\psi(\mathbf{r},t)$ is the complex,
dimensionless superconducting order parameter normalized with respect to $T_{\mathrm c}$.
We consider the static case where the external fields are absent, thus
the electro-magnetic potentials can be absorbed into the order parameter.
We include a small sixth order term in the GL free energy to ensure a
numerical stability.
The coefficient $\alpha=\alpha_{ \psi } (\mathbf{r})$ is a  functional
of the order parameter and includes the effects of long-range elastic potentials,
temperature $T$, and disorder:
\begin{equation}\label{eq.alpha}
      \alpha_{\psi} (\mathbf{r})=\alpha _{0}[T_{\mathrm c}(\mathbf{r})-T]+\int d\mathbf{r}%
      ^{\prime }U(\mathbf{r}-\mathbf{r}^{\prime })|\psi (\mathbf{r}^{\prime
},t)|^{2}\,,
\end{equation}
where $T_{\mathrm c}(\mathbf{r})=T_{\mathrm c}(1+\tau (\mathbf{r}))$ describes
quenched fluctuations ($\tau (\mathbf{r})\in \left[ -\lambda ;\lambda \right]
$) of $T_{\mathrm c}$ and $U(\mathbf{r-r^{\prime})}$ is the non-local kernel
generated by elasticity interactions.
The temperature $T$ is measured in units of superconducting
transition temperature $T_{\mathrm c}$,
and the unit of the length is the superconducting,
zero temperature coherence length $\xi$.

\subsection{Elastic interaction.}
To couple elasticity and superconductivity we utilize
the pressure dependence of $T_c(p)$ which is taken in a linearized form as
\begin{equation}\label{eq.Tcp}
T_c(p=p_0+\Delta p)=T_{c}(p_0)+\Delta p\frac{\partial T_c(p_0)}{\partial p}\,,
\end{equation}
where $\Delta p$ is the change in the internal pressure induced by elastic forces.
The latter is derived from the condition of the continuity of the deformation
fields ${\bf u}^{(s,p)}$ and the stress balance at the film-substrate interface
(the $z=0$ plane),
\[
\sigma _{iz}^\film(0,0,0)=\sigma _{iz}^\substr(0,0,0)\hspace{0.5cm}\textrm{for}\hspace{0.5cm}i=x,y,z\,,
\]%
and
\[
u_{i}^\film(0,0,0)=u_{i}^\substr(0,0,0)\hspace{0.5cm}\textrm{for}\hspace{0.5cm}i=x,y,z\,,
\]
and at the film-vacuum ($z=d$, $d$ is the thickness of the film) interface:
\begin{equation}
\sigma _{iz}^\film(0,0,d)=0\hspace{0.5cm}\textrm{for}\hspace{0.5cm}i=x,y,z\,.
\end{equation}
The superscripts $\film$ and $\substr$ refer to the film and to the substrate respectively.

These nine coupled equations are solved using a plain-wave Ansatz,
yielding the exact expressions for the deformation fields which, in their turn,
are used for calculating the internal pressure of the film.  The Fourier transform of the
pressure in the film then reads:
\begin{equation}\label{eq.press}
p^\film({\bf q})=\frac{1}{3d}\int_{0}^{d}dz\left( \partial _{x}u_{x}^\film+\partial
_{y}u_{y}^\film+\partial _{z}u_{z}^\film\right)\,.
\end{equation}

Using this expression together with the relation $\Delta p=-K\Delta V/V$
in Eq.~(\ref{eq.Tcp}) one finds the parameter $\alpha$
of the TDGL in the form given in (\ref{eq.alpha}).
We define the interaction potential $U(r)$ through its Fourier
transform $U_0\,{\cal K}(q)$, where $U_0=3K\Delta\alpha_L[\partial T_c(p_0)/{\partial p}]$
and ${\cal K}(q)$ is the scale free part of $p^\film({\bf q})/|\psi_q|^2$
[note, that $p^\film({\bf q})\propto |\psi_q|^2$].
Here $K$ is the bulk modulus, the linear thermal expansivity of the film is
$\Delta\alpha_L$, and the relative volume change due to the change
in pressure $\Delta p$ is $\Delta V/V$.

\subsection{Numerical simulation.}
The TGLE~(\ref{eq.tgle}) is solved on a fine discrete grid by
the numerical integration in real- and Fourier spaces
({\it quasi-spectral split-step method}), taking into account the
full non-local elastic potential.
We use periodic boundary conditions and $N=512^2$ grid points for a system
of the size $L^2=(200\xi)^2$.
The temperature is chosen as $T=0.8T_{\mathrm c}$ and the coupling constant of
$U_0/U_{\mathrm c}=2.23$ (in Figs.~2 and 3)
for a film of thickness $d=0.8\xi$.
The used GL parameters are $\alpha_0=4.37$, $\beta=1/2$, $\gamma=0.01$, and $\delta=0.1$.
The parameters of the elastic kernel are $\mu^{(s)}=0.5$, $\mu^{(p)}=5.0$
(shear moduli) and $\nu^{(s,p)}/(1-2\nu^{(s,p)})=1.6$
(modified Poisson numbers) -- see supplementary materials for details.

\section{Supplementary calculations}

\subsection{Construction of the elastic interaction potential $U$}

Here we derive $U\left( \mathbf{r}\right) $ for the problem of a thin
superconducting elastic film coupled to a rigid substrate with different elastic properties and lattice
mismatch. In the following, we use the superscripts $\film$ for all quantities
related to the film, and $\substr$ for the substrate.
Both materials are described by the three dimensional
displacement fields $\mathbf{u}^\film$/$\mathbf{u}^\substr$.
We start our consideration using a plain wave Ansatz for these fields:
\begin{eqnarray*}
u_{x}^\film &=&e^{\imath {q_{x}}x+\imath {q_{y}}y}\left(
e^{q_{z}z}A_{1,x}+A_{2,x}e^{-q_{z}z}+u_{0,x}^\film\right) \\
u_{y}^\film &=&e^{\imath {q_{x}}x+\imath {q_{y}}y}\left(
e^{q_{z}z}A_{1,y}+A_{2,y}e^{-q_{z}z}+u_{0,y}^\film\right) \\
u_{z}^\film &=&e^{\imath {q_{x}}x+\imath {q_{y}}y}\left(
e^{q_{z}z}A_{1,z}+A_{2,z}e^{-q_{z}z}\right) \\
u_{x}^\substr &=&C_{x}e^{{q_{z}}z+\imath {q_{x}}x+\imath {q_{y}}y} \\
u_{y}^\substr &=&C_{y}e^{{q_{z}}z+\imath {q_{x}}x+\imath {q_{y}}y} \\
u_{z}^\substr &=&C_{z}e^{{q_{z}}z+\imath {q_{x}}x+\imath {q_{y}}y}\,,
\end{eqnarray*}
where the misfit strain on the film is captured by the real exponential
terms of the film in $z$-direction.
$u_{0,x}^\film$ and $u_{0,y}^\film$ are special solutions of the two-dimensional problem due to superconductivity,
which we will discuss later.
In order to determine the nine free parameters $C_i$, $A_{1,i}$,
and $A_{2,i}$ we need nine equations.
First, we consider the interface film-vacuum at $z=d$,
where $d$ is the thickness of the film.
The stress boundary condition for a flat
surface demands: $\sigma _{xz}=\sigma _{yz}=\sigma _{zz}=0$.
So, for the film we get from Hooke's law
\begin{eqnarray*}
\sigma _{xz}^\film &=&\mu^\film\left( \partial _{z}u_{x}^\film+\left( 1-2\sigma
_{0}\right) \partial _{x}u_{z}^\film\right) \\
\sigma _{yz}^\film &=&\mu^\film\left( \partial _{z}u_{y}^\film+\left( 1-2\sigma
_{0}\right) \partial _{y}u_{z}^\film\right) \\
\sigma _{zz}^\film &=&2\mu^\film\left[ \left( 1+\nu^\film\right) \partial
_{z}u_{z}^\film+\nu^\film\left( \partial _{x}u_{x}^\film+\partial
_{y}u_{y}^\film\right) \right]\,,
\end{eqnarray*}
where the effect of a lattice mismatch of both materials is captured
by the $\sigma_{0}$-term.
In general, the lattice constants of both materials are different, i.e., the misfit parameter
$\eta=(a^\film-a^\substr)/a^\substr$
is non-zero, where $a^\film$ and $a^\substr$ are the lattice constants
of the film and substrate, respectively. This lattice mismatch leads to a compression of the film-lattice
at the interface, such that the lattice spacings match there.
We note, that this description not only applies to crystalline structures but also for amorphous materials,
where the ''lattice'' constants are averaged quantities.


The coefficients $\mu^\film$ and $\nu^\film$
are the shear modulus and the modified Poisson number of the film
[$\nu^\film=\tilde\nu^\film/(1-2\tilde\nu^\film)$, where $\tilde\nu^\film$ is the Poisson number].
In the following we use the fact, that all interface equations
are independent of the $x$ and $y$ coordinate and set $x=y=0$.
Therefore, the first three of the nine equations are
\begin{equation}
\sigma _{iz}^\film(0,0,d)=0\text{ for }i=x,y,z\,.
\end{equation}

Next, we consider the interface between film-substrate  at $z=0$,
for which we also need the expressions for the stress tensor elements
\begin{eqnarray*}
\sigma _{xz}^\substr &=&\mu^\substr\left( \partial _{z}u_{x}^\substr+\partial
_{x}u_{z}^\substr\right) \\
\sigma _{yz}^\substr &=&\mu^\substr\left( \partial _{z}u_{y}^\substr+\partial
_{y}u_{z}^\substr\right) \\
\sigma _{zz}^\substr &=&2\mu^\substr\left[ \left( 1+\nu^\substr\right) \partial
_{z}u_{z}^\substr+\nu^\substr\left( \partial _{x}u_{x}^\substr+\partial
_{y}u_{y}^\substr\right) \right]\,.
\end{eqnarray*}
At this interface we need to fulfill six more conditions, where
three of them are resulting from the stress balance
\[
\sigma _{iz}^\film(0,0,0)=\sigma _{iz}^\substr(0,0,0)\text{ for }i=x,y,z\,,
\]%
and the last three equations are the continuity equations for the displacement fields
\[
u_{i}^\film(0,0,0)=u_{i}^\substr(0,0,0)\text{ for }i=x,y,z\,.
\]
These nine equations can now be solved in order obtain the free parameters of our Ansatz.

The two special solutions for $u^\film$ in x- and y-direction, $u_{0,i}^\film$, already used in the Ansatz
for the deformation fields of the film are obtained through solution of the extended stress-strain balance
\begin{equation}\label{eq.couple}
\beta^{(s)}\Delta \mathbf{u}^{(s)}+\nabla(\nabla \mathbf{u}^{(s)})
=\gamma^{(s)} \nabla |\psi(\mathbf{r})|^2\,,
\end{equation}
which takes into account that the systems energy is the sum of
the superconducting energy and the mechanical deformation energy.
This relation introduces two phenomenological parameters: $\gamma^\film$ and $\beta^\film$ which together define the coupling constant of the elastic interaction and superconductivity which is determined explicitly below.
Eq.~(\ref{eq.couple}) can be solved and gives
\begin{equation}
u_{0,i}^\film=-\imath\gamma_s\frac{f(q)q_i}{(1+\beta_s)\mathbf{q}^2}\,,
\end{equation}
for $i=x,y$.

Using the solutions for all coefficients we find the internal pressure of the film
\begin{equation}\label{eq.press}
p^\film(q_x,q_y,q_z)=\frac{1}{3d}\int_{0}^{d}dz\left( \partial _{x}u_{x}^\film+\partial
_{y}u_{y}^\film+\partial _{z}u_{z}^\film\right)\,.
\end{equation}
This expression simplifies to $p^\film(q_z)$ if we use the homogeneous parametrization
$q_x=q_z\cos(\theta)$, $q_y=q_z\sin(\theta)$. Due to the special solution for $u^\film$ the pressure
$p^\film(q)$ is proportional to $|\psi_q|^2$, where $\psi_q$ are the Fourier components of the order parameter.

Now, we need to consider the connection of elasticity and superconductivity.
For that, we first expand $T_c$ in changes of (internal) pressure:
\begin{equation}
T_c(p=p_0+\Delta p)=T_{c}(p_0)+\Delta p\frac{\partial T_c(p_0)}{\partial p}\,.
\end{equation}
The pressure change $\Delta p$ is related to a volume change $\Delta V$ as $K=-V\partial p/\partial V$, where $K$ is the bulk modulus.
Therefore we can write
\begin{equation}
\Delta p=-K\frac{\Delta V}{V}=-3K\frac{\Delta L}{L}\propto -3K\alpha_L\,,
\end{equation}
where $\Delta L/L$ is the relative length change of the system and $\alpha_L$ the linear thermal expansivity of the film.

Using the above result in the pressure-expanded critical temperature, the
parameter $\alpha$ in the TGLE obtains thus the non-local form shown in the \textbf{Methods Summary} section:
\begin{equation}\label{eq.alpha}
\alpha_{ \psi } (\mathbf{r})=\alpha _{0}[T_{c}(\mathbf{r})-T]+\int d\mathbf{r}%
^{\prime }U(\mathbf{r}-\mathbf{r}^{\prime })|\psi (\mathbf{r}^{\prime
},t)|^{2}\,,
\end{equation}
where $T_{c}(\mathbf{r})$ includes also spatial variations of $T_c$ due to weak disorder.

If we define the scale free elastic kernel
\begin{equation}
{\cal K}(q)=\frac{(1+\beta^{(s)})}{\gamma^{(s)}} \frac{p^{(s)}(q)}{|\psi(\mathbf{r})|^2}\,,
\end{equation}
where the explicit form of the elastic kernel is quite involved and given in section~\ref{app.epot}. Finally, we write the Fourier transform of $U(\mathbf{r})$ as $U_0\,{\cal K}(q)$, where $U_0=3K\Delta\alpha_L[\partial T_c(p_0)/\partial p]$

\subsection{Numerical realization}
The TGLE equation
\begin{equation}\label{eq.tgle}
\partial _{t}\psi =\alpha\psi -\beta |\psi |^{2}\psi +\gamma \mathbf{\nabla }
^{2}\psi-\delta |\psi |^{4}\psi\,,
\end{equation}
is solved by a quasi-spectral split step method for periodic
boundary conditions on a two-dimensional square grid with
grid size of $N^{2}$ for a system size $L^{2}$.
{\it Split step} means, that for each time step the equation
is solved partially in real space and partially in Fourier space
(the diffusion part)~\cite{aranson_2002RMP}.
In the  first step we calculate
\begin{equation}
\psi_{ij} (t+\Delta t)=e^{\Delta t(\alpha-\beta |\psi_{ij}|^2)}|\psi_{ij}(t)|^2\,,
\end{equation}
where the non-local part of $\alpha$ is calculated by Fast-Fourier-Transformation (FFT) upfront.

Then, in the second step, we first Fourier transform $\psi_{ij} (t+\Delta t)$,
apply the diffusion kernel, and transform it back.
This way we avoid mostly all complications of the diffusion equation~\cite{aranson_2002RMP}.

In general the TGLE without long-range potential has only two stable solutions:
a homogeneous one and a striped phase.
Even if a long-range potential is present, the stability of these two solutions
is only destroyed under certain conditions which can be found out by linear stability
analysis (see next section). E.g., a Coulomb potential or even screened Coulomb potential cannot create
this kind of instability.
However, the elastic potential does, if the model parameters are chosen appropriately.
Another possible non-local potential which destroys the homogeneous/striped solutions
is a box-potential~\cite{pomeau+prl94} $U(\mathbf{r})=U_0\Theta(1-|\mathbf{r}|/a)$.

\subsection{Elastic kernel and simulation parameters}\label{app.epot}

\begin{table}\label{tab.param}
\caption{List of parameters for the elastic model.}
\begin{tabular}{ll}
$d$ & thickness of the film \\
$\mu ^{(s,p)}$ & shear modulus \\
$\sigma_{0}$ & deformation stress \\
$\nu^{(s,p)}$ & modified Poisson number \\
$U_{0}$ & potential strength \\
$\kappa $ & coupling constant \\
$\sigma_{ij}^{(s,p)}$ & stress tensor \\
$u_{i}^{(s,p)}$ & displacement fields \\
$A_{1,i},A_{2,i},C_{i}$ & 9 free variables for the elastic equations%
\end{tabular}
\end{table}

In order to determine whether the elastic long-range interactions lead to an instability of the homogeneous solution $\psi_0$ of the GL equation, we need to check the linear stability of the equation, by expanding the order parameter around $\psi_0$.
After this expansion and Fourier transformation, we find that the stability depends on the behavior of the function
\begin{equation}
S(q)=(2 U_0 {\cal K}(q)-1)(1-T/T_c)-\gamma q^2/9.38\,.
\end{equation}
We know that ${\cal K}(q=0)=0$ and ${\cal K}(q\to\infty)=1/3$ and therefore $S(q=0)<0$ (for $T<T_c$) and
$S(q)\to -\infty$ for $q\to\infty$.
If $S(q)$ changes its sign at intermediate $q$ the homogeneous solution becomes unstable, see Fig.~\ref{fig.stab}.

\setcounter{figure}{4}
\begin{figure}[tbh]
\includegraphics[width=0.9\columnwidth]{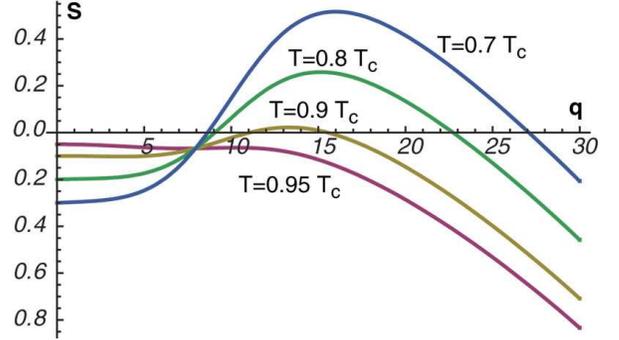}
\caption{{\bf Linear stability}. Plot of the stability finction $S(q)$ with elastic parameters
chosen as described in the text for different temperatures and $U_0=2.6 U_c$ (cf. inset of Fig.~4).}
\label{fig.stab}
\end{figure}

For completeness we write the scale free elastic kernel ${\cal K}$ explicitly, depending only on dimensionless material parameters:

\begin{widetext}
{\footnotesize
\begin{eqnarray}
{\cal K}(q)  & =&\left\{  \mu_{p}^{2}\left(  1+2\nu^\substr\right)  \left[  d~q(1+e^{4d~q}%
)\vartheta_{s}-(e^{4d~q}-1)\left(  2\vartheta_{s}+1\right)
\right]  \right. \nn \\
&& \left.  +d~q\left[  e^{4d~q}\mu_{s}\vartheta^\film\left(  2\mu^\substr%
+2\mu^\substr\nu^\substr\sigma_{0}-\mu^\film\vartheta^\film\right)  \right.
\right. \nn \\
&& \left.  ~~~\left.  -\mu^\film\vartheta^\film\left(  \mu^\film\vartheta^\film+
2\mu^\substr\left[  1+\nu^\substr+\nu^\film-\nu^\substr\sigma_{0}\right]  \right)
\right.  \right. \nn \\
&& \left.  ~~~\left.  -2e^{2d~q}\left(  {\mu^\substr}^{2}\left[  1+2\nu^\substr\right]
\left[  1+2\nu^\film\sigma_{0}\right]  -{\mu^\film}^{2}{\vartheta^\film}^{2}%
-\mu^\substr\mu^\film\vartheta^\film(\nu^\substr+\nu^\film-2\nu^\substr\sigma_{0})\right)
\right]  \right. \nn \\
&& \left.  +\mu^\film\left(  e^{d~q}-1\right)  ^{2}\left[  4\mu^\film\nu^\film\left(
e^{2d~q}-1\right)  \vartheta^\film\left[  \sigma_{0}-1\right]  \right.
\right. \nn \\
&& \left.  \left.  ~+\mu^\substr\left[  1+2\nu^\film\left(  2-2\vartheta%
^\film-3\sigma_{0}\right)  +2\nu^\substr\left[  2\vartheta^\film+1\right]
\left[  \sigma_{0}-1\right]  -2e^{d~q}\left(  \nu^\film\left(  2\nu^\substr\left[
\sigma_{0}-1\right]  +2\sigma_{0}-3\right)  -1\right)  \right.  \right.
\right. \nn \\
&& \left.  \left.  \left.  ~~~-e^{2d~q}\left(  3\vartheta^\film+2+2\nu
^\substr\left[  \sigma_{0}-1\right]  \left(  \nu^\film\left[  4\sigma_{0}-2\right]
-1\right)  \right)  \right]  \right]  \right. \nn \\
&& \left.  \left.  +4{\mu^\substr}^{2}\nu^\film e^{2d~q}\left[  1+2\nu^\substr\right]
\left[  4\sigma_{0}-3\right]  \sinh\left(  d~q\right)  \right\}  \right/ \nn \\
&& \left\{  3d~q\left\{  -{\mu^\film}^{2}\left(  e^{2d~q}-1\right)  ^{2}%
{\vartheta^\film}^{2}\right.  \right. \nn \\
&& \left.  \left.  +2\mu^\substr\mu^\film\left(  e^{2d~q}-1\right) \vartheta^\film
\left(  1+\nu^\substr+\nu^\film-\nu^\substr\sigma_{0}+e^{2d~q}\left[  1+\nu
^\substr\sigma_{0}\right]  \right)  \right.  \right. \nn \\
&& \left.  \left.  +{\mu^\substr}^{2}(1+2\nu^\substr)\left(  e^{4d~q}\vartheta%
^\film+2\nu^\film\left[  \sigma_{0}-1\right]  -1-2e^{2d~q}\left[  1+2\nu^\film%
\sigma_{0}\right]  \right)  \right\}  \right\}
\end{eqnarray}
}
\end{widetext}
with $\vartheta^\film\equiv 2\nu^\film\left[ \sigma_{0}-1\right] -1$
and parameters described in table~\ref{tab.param}.

For the simulations (cf. evolution of amplitude and phase in Figs.~2 and 3) we used the typical values: $T=0.8T_c$, $U_0=2.2U_c$, $\gamma=0.01$,
$\beta=0.5$, $d=0.8\xi_0$, $\mu^\film=0.5$, $\mu^\substr=5$ (the substrate is more rigid), $\nu^\film=\nu^\substr=1.6$, $\sigma_0=0$ (the lattice mismatch in the normal state is not relevant), $L=200\xi_0$ (linear dimension of the system), and $N=512$ (number of discrete grid points per dimension).


\begin{thebibliography}{99}
\bibitem{note2} Note, in particular, the useful Ehrenfest relation
$\partial \log{T_{\mathrm c}}/\partial p \simeq\alpha_{\mathrm {\scriptscriptstyle L}}/
\Delta C_{\mathrm p}$, where $\Delta C_{\mathrm p}$
is the change in the specific heat at the
superconductor transition.
\end{thebibliography}
\end{document}